\documentclass[12pt]{article}
\usepackage{axodraw}
\usepackage{pstricks}
\title{Electromagnetic form factors of hadrons in the time-like region}
\author{O.D. Dalkarov, P.A. Khakhulin, A.Yu. Voronin}
\date{Lebedev Physical Institute, Moscow, Russia}

\begin{document}
\maketitle
\begin{abstract}
Hadron electromagnetic form factor in the time-like region at the boundary of the physical region is considered. The energy behaviour of the form factor is shown to be dominantely determined by the strong hadron-antihadron interaction. The scattering lengths for $p\bar{p}$, $\Lambda\bar{\Lambda}$, $\Lambda\bar{\Sigma}^0 (\bar{\Lambda}{\Sigma}^0)$ and ${\Sigma}^0\bar{\Sigma}^0$ are evaluated. The experiments to extract information on the nearthreshold $B\bar{B}$ interaction by using of hadron form factor properties are proposed. 
\end{abstract}

\section{Introduction}
The main goal to study electromagnetic form factors of hadrons is an obtaining of information about structure of these particles. The most complete data on the behaviour of the form factor as a function of four-momentum transeferred is obtained for pion and nucleon. To investigate form factor of hadron $(h)$ two reactions are used: the reaction of elastic scattering of electrons by hadron $eh \to eh$ (so-called space-like region of the four-momentum transferred) and the reaction of hadron-antihadron pair production in the electron-positron annihilation $e^{+}e^{-} \to h\bar{h}$ or inverse reaction (time-like region).

High precision data on the electromagnetic form factor of a proton in the time-like region firstly became available due to good quality experiment PS--170 performed at LEAR (CERN) and recent BaBar one, where the reactions $p\bar{p} \to \Lambda\bar{\Lambda}, \Lambda\bar{\Sigma}^0, {\Sigma}^0\bar{\Sigma}^0$ were reported~\cite{1}. Also BES collaboration~\cite{2} has observed the nearthreshold enhancement of the $\bar{p}$ - system in the $J/\psi \to Kp$ decay. These data reveal a feature which principally differs behaviour of the electromagnetic baryon form factor in the time-like region from that of a pion. The form factor drops enormously quickly just near threshold (by the factor of approximately two from the threshold to $3.6 \:{\rm GeV}^2$ for proton one i.e. in the region of c.m. relative momentum of the order $100\:{\rm MeV}/c$).

To describe the behaviour of the form factor different vector dominance models (VDM) are used. These models successfully reproduce the behaviour of the pion form factor both in space-like and time-like regions~\cite{3} and the behaviour of the nucleon (proton and neutron) form factors in the space-like regions~\cite{4}, but they fail in description of the proton form factor in the time-like region. Also several explanations of the energy behaviour for proton form factor were considered after obtaining of the experimental data (see~\cite{5} and references therein).

Another approach was proposed~\cite{6,7} to understand physics of the behaviour of the baryon electromagnetic form factor in the time-like region just near threshold. This approach takes into account final state interaction (interaction in the baryon-antibaryon system) as dominant physical reason giving energy form factor behaviour and its value near $B\bar{B}$ threshold. In this model, the form factor is separated into two parts according to different physical processes. Form factor is presented as a product of a factor corresponding to singularities of transition  amplitude lying far from $B\bar{B}$ threshold and a factor reflecting strong final state interaction. The energy dependence of the form factor is given by the latter factor. Moreover, the behaviour of the form factor appears to be directly connected to other observables in the $B\bar{B}$ system. For instance, it is possible to extract a value of imaginary part of baryon-antibaryon scattering length using the momentum dependence of the form factor just near $B\bar{B}$ threshold. Note that the consideration wich was proposed in~\cite{6,7} is a natural consequence of the main features of quasinuclear model for low energy baryon-antibaryon interaction~\cite{8} in which observed enhancement near threshold was predicted long time before experimental indications.

This approach predicts peculiar behaviour of the proton form factor in the time-like region and is able to reproduce experimental data. Moreover, it turns to be possible to predict the behaviour of the neutron form factor. These predictions differ sufficiently from the predictions of standart VDM. The neutron form factor can not exceed the proton one, whereas all vector dominance models obtain the neutron form factor five-ten times more than the proton one.

Established connection of the behaviour of the hadron form factor with the interaction in the hadron-antihadron system gives us unique possibility to investigate these systems. The main advantage of the electron-positron production of baryon-antibaryon pairs is that there is no initial state interaction and transition mechanism is well-determined. However, even existing facilities allows us to obtain such an information from the ''form-factor'' experiments with $e^{+}e^{-}$ beams (for instance for systems with hidden strangeness like $Y\bar{Y}$, or with hidden charm and beauty like $D\bar{D}$, $B\bar{B}$, etc.) We discuss here existing data on the $Y\bar{Y}$ production.

There is an additional group of experiments with $e^{+}e^{-}$ beams wich can give more information about hadron-antihadron interaction in final state. For a case of nucleon-antinucleon we can call experiments on investigation of electron-positron annihilation into multipion systems which are dominant modes of nucleon-antinucleon annihilation. In this reactions, system nucleon-antinucleon appears as an intermediate state. Hence quantum numbers of the $N\bar{N}$ system are difinitely fixed, because nucleon and antinucleon are created from electron-positron pair through a photon what gives photon quantum numbers to the nucleon-antinucleon system. Even or odd amount of pions in the final state fixes isospin of a system. So this kind of experiments allows us prepare the nucleon-antinucleon system in a state with definite quantum numbers, that is practically difficult in any other experiment. This fact makes investigation proposed here very fruitful. For example, the deep-bump structure~\cite{9} obeserved in the electron-positron annihilation into six pions near nucleon-antinucleon threshold was interpretated~\cite{7} as an expected manifistation of Green function zero for interacting $N$ and $\bar{N}$ in the intermediate state~\cite{10}. From this data the existence of a new narrow vector $N\bar{N}$ state was predicted~\cite{7}. With $\bar{p}$ beams the direct observation  of new subthreshold $N\bar{N}$ state is possible with nuclear targets, but even in deuteron case an extracting of parameters of such kind of state needs special theoretical considerations due to cancellation for non-adiabatic and off-mass shell effects~\cite{11}.

The main advantage of the approach presented in~\cite{6,7} and here is its ability to describe simultaneously the properties of nucleon-antinucleon interaction, electromagnetic form factor of the nucleon in the time-like region, and multipion electron-positron annihilation near $N\bar{N}$ threshold. These features could be found in other hadron-antihadron systems considered here.

The article is organized the following way. Section 2 is devoted to the general properties of the form factor in the time-like region. The manifestation of these properties on the example of the nucleon form factor is considered in Secton 3. In Section 4 we discuss form factor properties of other hadrons. Conclusion and proposals for experiments are presented in Secton 5.

\section{General properties of the form factor}
In this Section we investigate general properties of the hadron form factor in the time-like region near the boundary of the physical region. For different hadrons we can have different number of form factors, because it depends on spin of hadron. Hence as a concrete example we will write formulas for a nucleon (proton and neutron). But all physical results can be trivially generalized for other hadrons what will be done in Section 4.

The form factor of the nulceon ($N$) in the time-like region is determined from the reaction of $e^{+}e^{-}$-annihilation into nucleon-antinucleon pair $e^{+}e^{-} \to N\bar{N}$ or vice versa. This is so-called s--channel in contrast to t--channel corresponding to $eN \to eN$ scattering or to the nucleon form factor in the space-like region.

The differential cross section $d\sigma/d\Omega$ of the reaction $p\bar{p} \to e^{+}e^{-}$ is connected to the form factor of the proton in the vicinity of $p\bar{p}$ threshold by the formula
\begin{equation}
	\frac{d\sigma}{d\Omega} = \frac{\alpha^2}{32kE}\left(|G_{M}|^2 (1+\cos^2\theta) + \frac{{4M_p}^2}{s}|G_E|^2 \sin^2\theta\right),
\end{equation}
here $k$ and $E$ is center-of-mass momentum and energy in the $p\bar{p}$ system, $\theta$ is angle in c.m.s., $\alpha$ is the fine structure constant, $M_p$ is the proton mass. $G_E$ and $G_M$ are electric and magnetic form factors of the proton correspondingly. They are connected to Pauli form factors $F_1$ and $F_2$:
\begin{equation}
	G_E = F_1+F_2, \;\;\; G_M = F_1-\frac{q^2}{4{M_N}^2}F_2,
\end{equation}
here $q$ is four-momentum transferred, which is equal to $t$ in the center-of-mass system of the reaction $eN \to eN$ and to $s$ in the reaction $e^{+}e^{-} \to N\bar{N}$. Threshold of the latter reaction corresponds to $q^2 = -4{M_N}^2$ ($M_N$ is the nucleon mass).

At the $p\bar{p}$ threshold $G_E$ and $G_M$ are equal and for simplicity hereafter they are taken to be equal in the kinetic energy region of few tens of MeV near the threshold.

\begin{figure}[h]
\begin{center}
\fcolorbox{white}{white}{
  \begin{picture}(398,94) (142,-111)
    \SetWidth{0.5}
    \SetColor{Black}
    \GBox(239,-97)(450,-35){1.000}
    \Vertex(375,-66){5.00}
    \Vertex(270,-66){2.00}
    \ArrowLine(270,-66)(165,-36)
    \ArrowLine(165,-96)(270,-66)
    \ArrowLine(480,-36)(375,-66)
    \ArrowLine(375,-66)(480,-96)
    \Photon(270,-66)(375,-66){7.5}{5}
    \GOval(436,-66)(17,6)(0){0.350}
    \Text(370,-87)[lb]{\Large{\Black{${\scriptstyle G_0}$}}}
    \Text(322,-55)[lb]{\Large{\Black{${\scriptstyle \gamma}$}}}
    \Text(485,-39)[lb]{\Large{\Black{${\scriptstyle \bar{N}}$}}}
    \Text(485,-99)[lb]{\Large{\Black{${\scriptstyle N}$}}}
    \Text(150,-39)[lb]{\Large{\Black{${\scriptstyle e^{+}}$}}}
    \Text(150,-99)[lb]{\Large{\Black{${\scriptstyle e^{-}}$}}}
    \Text(344,-111)[lb]{\Large{\Black{${\scriptstyle G}$}}}
  \end{picture}
}
\end{center}
\caption{The diagram corresponding to electron-positron annihilation into the nucleon-antinucleon system.}
\label{diagram}
\end{figure}
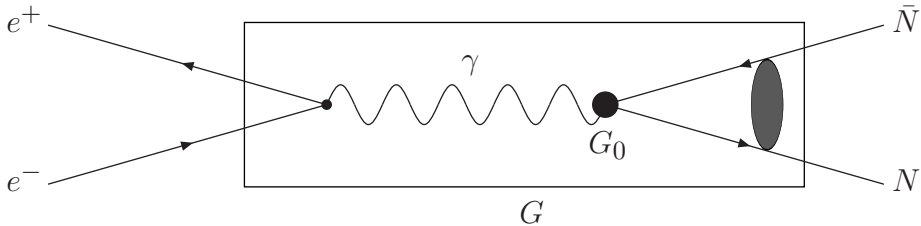

Before doing any calculations we can make some conclusions about nucleon form factor near threshold. Let's consider a diagram corresponding to the process $e^{+}e^{-} \to N\bar{N}$ (Fig.~\ref{diagram}). Grey block in this diagram presents the final state interaction in the system $N\bar{N}$. We know that this interaction is very strong and even can produce bound states in the $N\bar{N}$ system~\cite{8}. Left part of the diagram corresponds to the transition amplitude from $e^{+}e^{-}$ pair into $N\bar{N}$ pair. Black circle in this transition amplitude denotes a connection between a photon and $N\bar{N}$ pair, which can be realized for example by vector mesons($\rho$ or $\omega$).

This transition amplitude without final state interaction corresponds to the s--channel diagrams and therefore has no t--singularities. So this amplitude in $r$-representation is proportional to $\delta$-function of $r$. In the first Born approximation on the fine structure constant this fact allows us rewrite immediately the expression for the transition amplitude $e^{+}e^{-} \to N\bar{N}$ and therefore for the form factor $G$ in the following way~\cite{6}
\begin{equation}
\label{ff}
	G = G_0|\Psi(0)|,
\end{equation}
where $\Psi(0)$ is the $N\bar{N}$ wave function in the origin.

In this model the form factor is separated into two parts: $G_0$ and $|\Psi(0)|$. The former corresponds to singularities far from $N\bar{N}$ threshold (for instance, to a connection of the photon to the nucleon through $\rho$- or $\omega$-exchanges, as it can be written in usual vector dominance models). The latter reflects strong final state interaction. Main dependence on energy of the form factor in the nearthreshold region will be given by the second factor in the formula (\ref{ff}). Since we are interested in the kinetic energy region of few tens of MeV in the $N\bar{N}$ center of mass system, the first factor $G_0$ is practically constant. So to determine energy dependence of the form factor it is necessary to investigate $\Psi(0)$.

The wave function of the $N\bar{N}$ system in the origin is directly connected to the Jost function $f(k)$ of the $N\bar{N}$ system~\cite{12}:
\begin{equation}
	\Psi(0) = \frac1{f(-k)},
\end{equation}
which can be rewritten in the form 
\begin{equation}
	f(k) = \frac{e^{i\delta(k)}}{\tau(k^2)}.
\end{equation}
Here $\delta(k)$ is the phase shift of the $N\bar{N}$ scattering and $\tau(k^2)$ is an even function of $k$ ($k$ is the momentum in c.m.s.).

When we are working in the region close to the $N\bar{N}$ threshold we can neglect the $k^2$-dependence of $\tau(k^2)$ and use scattering length approximation for the phase shift
\begin{equation}
	\delta(k) = -\alpha k,
\end{equation}
where $\alpha$ is the $N\bar{N}$ triplet scattering length.

So in the nearthreshold region the behaviour of the form factor is determined by the formula~\cite{7}
\begin{equation}
\label{Gexp}
	G = C e^{{\rm Im}\:\alpha\:k}.
\end{equation}

Exponential factor here provides us very sharp peak in the form factor behaviour near threshold. Moreover we can directly investigate the properties of the final system and determine a value of the scattering length. Just near threshold we can expand the exponent i.e.
\begin{equation}
	G = C(1+{\rm Im}\:\alpha\:k).
\end{equation}
Therefore in high precision experiment we can observe linear $k$-term in the behaviour of the form factor. This linear $k$-dependence is a direct manifistation of the threshold and final state interaction.

For very small $k$ the role of the Coulomb corrections is a special interest. In this case asymptotic behaviour of $\Psi$ function at large distances is given by
\begin{equation}
	\Psi \propto \frac1{r} \sin\left(kr+\frac{\ln(2kr)}{ka_c} + {\delta}_c +\delta \right)
\end{equation}
where $a_c = 2\hbar^2/(M_p\:e^2)$ is Coulomb length unit and ${\delta}_c = \arg \Gamma (1-i/(ka_c))$ is Coulomb phase shift, $M_p$ is a proton mass. Additional phase shift $\delta$ is caused by the short-range interaction. 

As it will be seen from Fig.~\ref{protonff} Coulomb corrections take place only at very small relative momentum $(p < 50\: {\rm MeV}/c)$, where $\bar{p}$ is deaccelerated due to electromagnetic loses and finally captured and annihilates from the protonium state.
\section{Comparison with experimental data}
\begin{figure}[ht]
\input{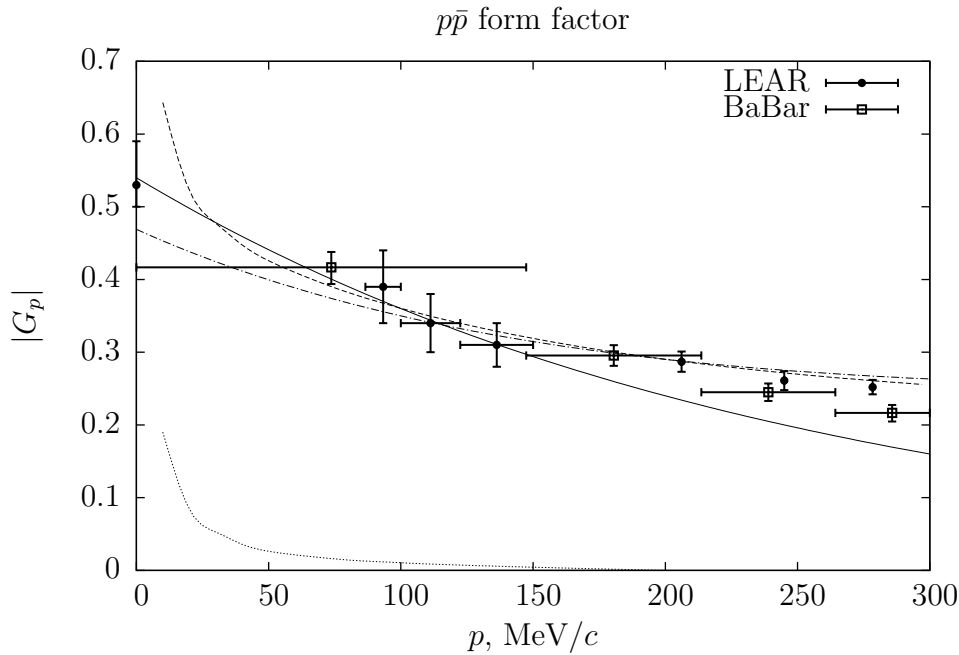}
\caption{Proton electromagnetic form factor in the time-like region. Experimental data are taken from~\cite{1}. Solid line is calculation in this work using formula (\ref{Gexp}) with ${\rm Im}\:\alpha = -0.8\:{\rm fm}$. Dotted and dash-dotted line are calculations using optical model~\cite{14} with and without Coulomb corrections, the lowest curve is the difference between calculations with and without Coulomb corrections.}
\label{protonff}
\end{figure}

Very precise experimental data on the proton form factor just near threshold were obtained in the LEAR and BaBar experiment~\cite{1}. These data with other ones are presented in Fig.~\ref{protonff}. We see sharp peak at small relative momenta with decreasing of the form factor in two times.

Note that in the approach presented here before exact calculation we can reproduce this energy behaviour. Namely we can extract value of imaginary part of scattering length from the data on the protonium ($p\bar{p}$ Coulomb atom) using the experimental values of shifts and widths of atomic S--levels~\cite{13}. In fact, we need scattering length corresponding to triplet state because final $p\bar{p}$ system in diagram in Fig.~\ref{diagram} is produced only in state with quantum numbers of a photon. This imaginary part of the scattering length from protonium data is equal to ${\rm Im}\:\alpha({}^3 S_1) = -0.8\:{\rm fm}$. The calculations of the form factor by using formula~(\ref{Gexp}) with this value of ${\rm Im}\:\alpha$ and $C=0.54$ are persented in Fig.~\ref{protonff} by solid line (this line crosses four first LEAR experimental points). The calculations using optical model~\cite{14} (with and without Coulomb corrections) are presented in Fig.~\ref{protonff} by dotted and dash--dotted lines (with normalization at $p=200\:{\rm MeV}/c$). Imaginary part of the scattering length corresponding to this curve is ${\rm Im}\:\alpha = -0.70\:{\rm fm}$.

By using of experimental data on the proton form factor we can predict a value of the neutron form factor just near threshold. For this let's write the definition of the nucleon form factor in terms of the isoscalar (isospin is equal to $0$) $G(0)$ and isovector $G(1)$ (isospin 1) form factors:
\begin{equation}
	G_p = |G(0)+G(1)|, \; G_n = |G(0)-G(1)|.
\end{equation}
We can use this decomposition for both isospins and rewrite our formulas in terms of input form factors $G_0$ and wave functions of final state in pure isospin states:
\begin{equation}
	G_p = |G_0(0)\Psi(0) + G_0(1)\Psi(0)|, \; G_n = |G_0(0)\Psi(0) - G_0(1)\Psi(0)|.
\end{equation}

Depending on the relative values of $G_0(0)$ and $G_0(1)$ there a two possible situations.

\begin{figure}[ht]
\input{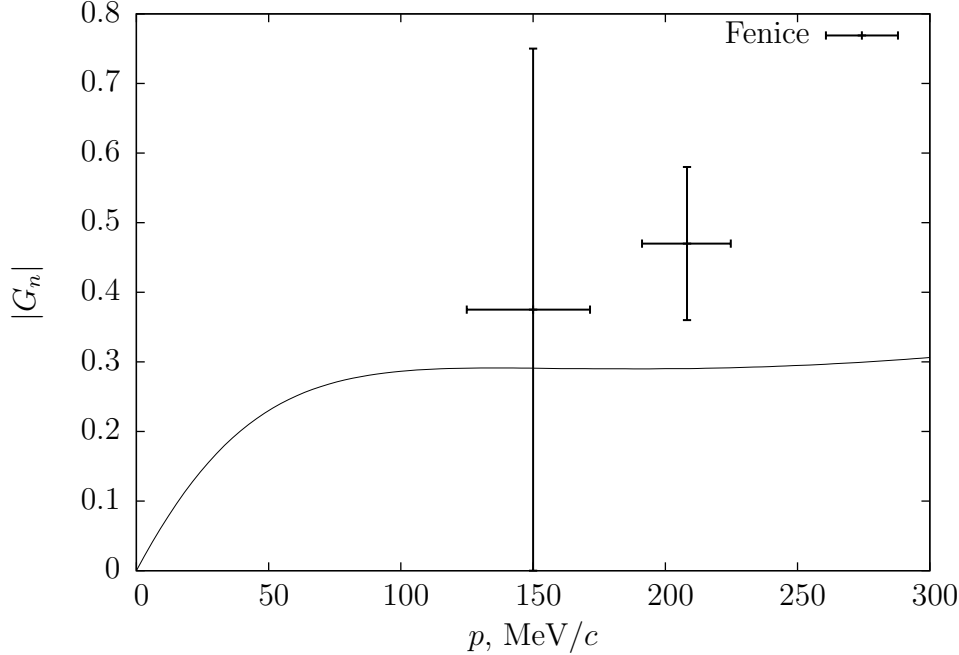}
\caption{Neutron form factor in the time-like region. Experimental data from~\cite{9}. Solid line is our calculation using optical model~\cite{14} with normalization at $p = 200\:{\rm MeV}/c$.}
\label{neutronff}
\end{figure}

First, $G_0(0) \approx G_0(1)$, i.e. there is no sufficient difference between isoscalar and isovector input form factors. In this case in the previous formulas we can take out of brackets the common factor $G_0$. We see immediately that neutron form factor $G_n$ is equal to or less than proton one. Note that in this case the energy behaviour of the neutron form factor can be different from the proton one. Just near threshold it can be practically constant or even decreasing function of the energy. The calculations of the neutron form factor using optical model~\cite{14} in this case are presented in Fig.~\ref{neutronff}. Note that the close predictions were done in coupled channels model in ref.~\cite{7}. Preliminary experimental data~\cite{9} indicates on realization namely this possibility.

Second, one of the input form factors is dominant, for instance, $G_0(1) \gg G_0(0)$. This situation seems to be probable if we believe that these form factors are determined by $\rho$- and $\omega$- mesons correspondingly. In this case we have dominance of $G_0(1)$ because $\rho$-meson has a product of coupling constants with nucleon and photon larger than that for $\omega$-meson. So we can neglect $G_0(0)$ contributions to the nucleon form factor and we obtain that proton and neutron form factors are approximately equal.

Therfore, in our model the neutron form factor does not exceed proton one in any case. This conclusion directly contradicts with predictions of the VDMs which gives neutron form factor five or even ten times more than proton one.

\section{Form factors of other hadrons.}
It is clearly seen that consideration presented above can be directly applied to investigation of the form factor of any hadron in time-like region.

Trivial generalization can be done for other baryons, first of all for baryons with strangeness. Closest threshold to the $N\bar{N}$ one is a threshold of $\Lambda \bar{\Lambda}$ production in $e^{+}e^{-}$-annihilation.

Let's estimate value of the lambda form factor near $\Lambda \bar{\Lambda}$ threshold. If we consider the same ideology of $e^{+}e^{-}$ transition through a vector meson into $\Lambda \bar{\Lambda}$ pair with consequent final state interaction, we obtain the following. A contribution of $\rho$-meson seems to be dominant. The value of $G_0$ for lambda is proportional to the coupling constants of $\rho$ meson with photon and $\rho$-meson with lambda. The former is known from the experiment. The latter can be estimated from the SU(3)-relations. Both are of the same order as for a nucleon. Final state interaction in $\Lambda \bar{\Lambda}$ system according to the existing approaches is approximately the same as in case of $N\bar{N}$. Hence there are no reasons for the lambda form factor to be strongly suppressed as compared to nucleon one and we expect them to have the same order of magnitude. Moreover the final state interaction in the $\Lambda \bar{\Lambda}$ system in some sense is even more clear than in the case of nucleon-antinucleon interaction, because here we are dealing with pure isospin ($I=0$) state.

New data on $\Lambda \bar{\Lambda}$, $\Sigma^0 \bar{\Sigma}^0$ and $\Lambda \bar{\Sigma}^0$ form factors became available recently due to experiments by BaBar group. In the work~\cite{1} they introduce effective form factor of a hadron as follows
\begin{equation}
	{|F(m)|}^2 = \frac{2\tau {|G_M(m)|}^2+{|G_E(m)|}^2}{2\tau + 1},
\end{equation}
where $\tau = m^2/4{m_B}^2$, $m$ is mass of hadronic system and $m_B$ is baryon mass. Since data on the ratio $|G_E|/|G_M|$ is imprecise(for $\Lambda\bar{\Lambda}$, see~\cite{1}) or even does not exist, hereafter we will assume that $|F|=|G|$ and then it becomes possible to estimate imaginary parts of scattering lengths in these systems. Experimental data and fitted exponential curves $|F| \approx C\exp({\rm Im}\:\alpha\:k)$ are presented below in Fig.~\ref{baryef}.

\begin{figure}[ht]
\input{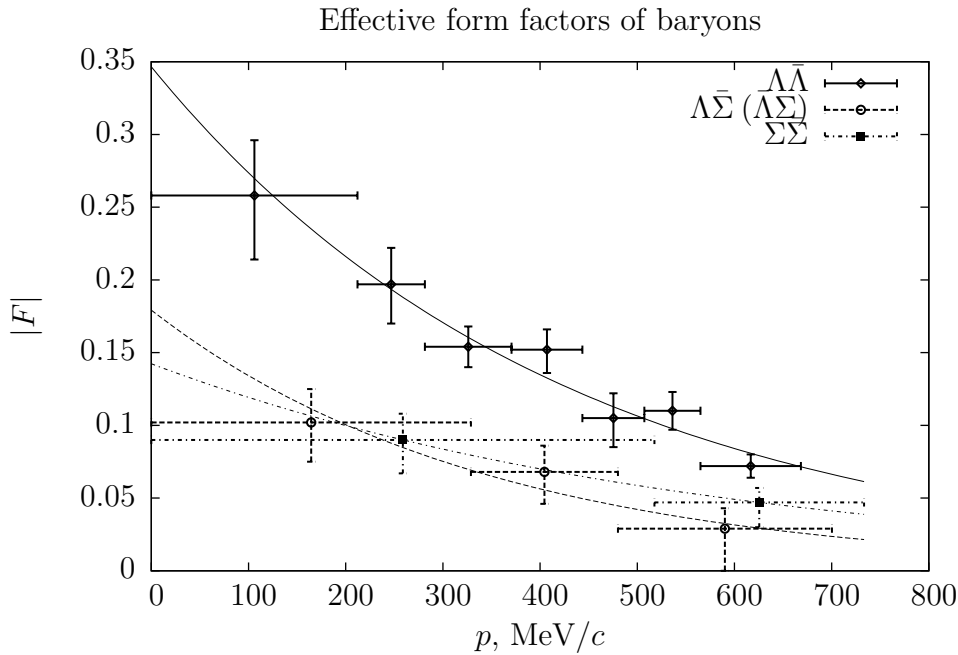}
\caption{Baryon effective form factors. Experimental data are taken from BaBar experiment~\cite{1}. Solid, dotted and dash-dotted lines are $\Lambda\bar{\Lambda}$, $\Lambda\bar{\Sigma}^0 (\bar{\Lambda}\Sigma^0)$ and $\Sigma^0\bar{\Sigma}^0 \;$ fitted exponential curves correspondingly.}
\label{baryef}
\end{figure}

Values of scattering lengths extracted from these data are presented in Table~\ref{scatlen}.

%\begin{table}[ht]
%\begin{center}
%\begin{tabular}{|c|c|c|}
%\hline
% & C & ${\rm Im}\:\alpha$, fm \\
%\hline
%\hline
%$\;\Lambda\bar{\Lambda}\;$ & $\;\;\;\; 0.35\;\;\;\;$ & $\;\;\; -0.47\;\;\;$ \\
%\hline
%$\;\Lambda\bar{\Sigma}^0 (\bar{\Lambda}\Sigma^0)\;$ & $\;\;\;\; 0.14\;\;\;\;$ & $\;\;\; -0.35\;\;\;$ \\
%\hline
%$\;\Sigma^0\bar{\Sigma}^0\;$ & $\;\;\;\; 0.18\;\;\;\;$ & $\;\;\; -0.57\;\;\;$ \\
%\hline
%\end{tabular}
%\end{center}
%\caption{Scattering lengths in $B\bar{B}$ system.}
%\label{scatlen}
%\end{table}

\begin{table}[ht]
\begin{center}
\begin{tabular}{|c|c|c|}
\hline
 & C & ${\rm Im}\:\alpha, {\rm fm}$ \\
\hline
$\;\Lambda\bar{\Lambda}\;$ & $\;\;\;\; 0.35\;\;\;\;$ & $\;\;\; -0.47\;\;\;$ \\
\hline
$\;\Lambda\bar{\Sigma}^0 (\bar{\Lambda}\Sigma^0)\;$ & $\;\;\;\; 0.18\;\;\;\;$ & $\;\;\; -0.57\;\;\;$ \\
\hline
$\;\Sigma^0\bar{\Sigma}^0\;$ & $\;\;\;\; 0.14\;\;\;\;$ & $\;\;\; -0.35\;\;\;$ \\
\hline
\end{tabular}
\end{center}
\caption{Scattering lengths in $B\bar{B}$ system.}
\label{scatlen}
\end{table}

Due to the same reason an investigation of the reactions of $e^{+}e^{-}$-annihilation near other $Y \bar{Y}$ thresholds will give unique information about these systems.

Absolutely analogous ideology can be applied to study meson-antimeson and other baryon-antibaryon systems (including systems with hidden charm and beauty). It will be very interesting to measure form factors of $D$-, $F$- mesons, $B^{+}\bar{B}^{+}, B^{-}\bar{B}^{-}$ - baryons, etc.

Note that very sharp behaviour can not be seen in the pion form factor, because $\pi^{+}\pi^{-}$ system has only elastic scattering and has no absorbtion at the threshold.

\section{Conclusion}
The analysis of the experimental data shows that the behaviour of the electromagnetic form factor of a hadron is mostly determined by the interaction of the hadron-antihadron in the final state. Therefore the measurements of the form factor properties can serve as a very fruitful source of information about hadron-antihadron interaction, especially in situation when direct investigation of this interaction is implossible.

To obtain more elaborate information about hadron-antihadron interaction the following experiments with electron-positron annihilation are desirable:
\begin{itemize}
\item Precise measurements of the proton and neutron form factors in the time-like region just near threshold of the reaction $e^{+}e^{-} \to N\bar{N}$ give us opportunity of high quality determination of $N\bar{N}$ scattering parameters.
\item Further investigation of the strange and charm particles electromagnetic form factors in the time-like region just near threshold. Especially for non-relativistic region of relative momentum less than few hundreds of ${\rm MeV}/c$.
\item There is a possibility to discover a ${}^3 S_1{}$ quasinuclear bound states in the $B\bar{B}$ system, which can manifest themselves as a heavy vector meson. To do it, the experiment to measure proton form factor near $B\bar{B}$ threshold is desirable, because these states will manifest themselves as the bumps in the form factor behaviour ($e^{+}e^{-} \to B\bar{B} \to p\bar{p}$). 
\item Bound states with photon quantum numbers in hadron-antihadron systems will manifest themselves also as a deep-bump structure in the electron-positron transition into main annihilation channels of these systems. In particular it will be interesting to search for phenomena connected with such a vector meson state in the $\Lambda\bar{\Lambda}$  system near threshold in the reaction $e^{+}e^{-} \to K\bar{K}4\pi$ by the analogy with deep-bump structure in 6$\pi$ annihilation channel near $N\bar{N}$ threshold.
\item To determine interaction in the systems with hidden new quantum numbers, the experiments on precise measurement of the cross sections $e^{+}e^{-} \to K\bar{K}, D\bar{D}, F\bar{F}, B^{+}\bar{B}^{+}, B^{-}\bar{B}^{-}$ etc near corresponding thresholds can be very informative.
\end{itemize}

In conclusion let's emphasize once more that in the frame work of the approach which was proposed in~\cite{6,7} and here just near each hadron-antihadron threshold it will be very interesting to measure both cross section of electron-positron transition into hadron-antihadron pair and cross sections of electron-positron transition into systems corresponding to main modes of hadron-antihadron annihilation.

\section*{Acknowledgement}
We would like to thank J. Carbonell for providing us with the program of computating optical potential.

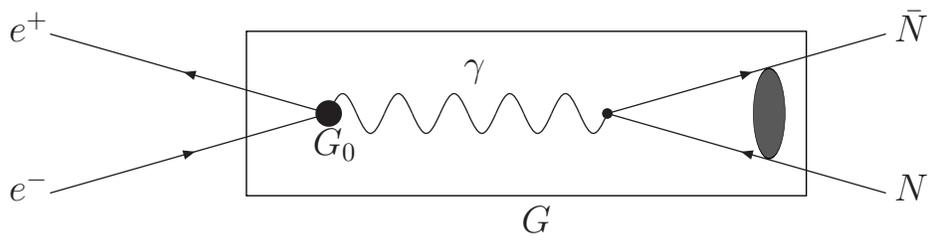
\end{document}